\documentclass[aps,prb,showpacs,twocolumn]{revtex4}  

\usepackage{graphicx}
\usepackage{amssymb}                

\def\newblock{\hskip .11em plus .33em minus .07em}
\def\gapx{\lower 2pt \hbox{$\buildrel>\over{\scriptstyle{\sim}}$\ }}
\def\lapx{\lower 2pt \hbox{$\buildrel<\over{\scriptstyle{\sim}}$\ }}
\def\he4{$^4$He}
\def\paraH2{{\it p}-H$_2$}
\def\orthoD2{{\it o}-D$_2$}
\def\Am2{\AA$^{-2}$}

\begin{document}

\title{Superfluid response  of 2D parahydrogen clusters in confinement}
\author{Saheed Idowu and Massimo Boninsegni} 
\affiliation{Department of Physics, University of Alberta, Edmonton, 
    Alberta, Canada T6G 2E7}
\date{\today}

\begin{abstract} 

We study by computer simulations the effect of confinement on the superfluid properties of  small two-dimensional (2D) parahydrogen clusters. For clusters of fewer than twenty molecules, the superfluid response in the low temperature limit is found to remain comparable in magnitude to that of 
free clusters, within  a rather wide range of depth and size  of the confining well. The resilience of the superfluid response is attributable to the ``supersolid" character of these clusters.  We investigate the possibility of establishing a bulk 2D superfluid ``cluster crystal" phase of \paraH2, in which a global superfluid response would arise from tunnelling of molecules across adjacent unit cells. The computed  energetics suggests that for clusters of about ten molecules, such a phase may be thermodynamically stable against the formation of the equilibrium insulating crystal, for values of the cluster crystal lattice constant possibly allowing tunnelling across adjacent unit cells.

\pacs{02.70.Ss,67.40.Db,67.70.+n,68.43.-h.} 
\end{abstract}

\maketitle

\section{INTRODUCTION}
The observation of the putative superfluid phase\cite{ginzburg} of condensed parahydrogen (\paraH2) has so far been prevented by its strong tendency to crystallize at low temperature, even in reduced dimensions.\cite{bon04,bon13} There exists, however, experimental\cite{vilesov} evidence that finite clusters of \paraH2 remain liquidlike down to temperatures much lower than the bulk crystallization temperature, conceivably allowing one to probe their predicted superfluid behaviour, expected to manifest itself at a temperature $T\sim 1$ K, for clusters of thirty molecules or less.\cite{sindzingre,MB1,MB2} The question thus arises of whether one might be able to observe a {\it macroscopic} superfluid response in a network of interconnected superfluid clusters, in which global phase coherence could be established by tunnelling of molecules across adjacent clusters. This is, in a sense, analogous to the physics of the recently proposed supersolid phase of soft core bosons in 2D.\cite{saccani,review,amj}  
\\ \indent
The above scenario could be realized experimentally, for example, by adsorbing \paraH2 inside a porous material such as vycor. A second avenue, perhaps affording greater control, exploits the predicted superfluid response of  \paraH2 clusters confined to quasi 2D.\cite{gordillo,saheed} 
The idea is that of fashioning a planar substrate capable of adsorbing \paraH2 molecules at specific sites, arranged on a regular triangular lattice, each designed to accommodate a number of molecules corresponding to a strong superfluid response at low temperature  (i.e., around twenty\cite{saheed}). The distance between nearest-neighboring clusters should be chosen to render such a
cluster crystal energetically favorable with respect to the formation of the ordinary uniform, non-superfluid crystalline phase (with just one molecule per unit cell),  while concurrently allowing tunnelling of molecules across nearest neighboring wells, each one acting in a sense like a superfluid quantum dot. Whether all of these conditions can be met is not {\it a priori} obvious, and furnishing a quantitative answer is the goal of this work.
\\ \indent
Besides the energetics, a second aspect to assess quantitatively is the effect of confinement on the superfluid properties of the individual clusters, which, besides being obviously crucial to the goal of stabilizing the bulk superfluid phase described above,  is of interest in its own right, and might be probed experimentally, for example by trapping small clusters at  adsorption sites of corrugated substrates.\cite{yang}
On the one hand,  spatial confinement is expected to bring about a reduction of the superfluid response of a structureless, liquidlike droplet, owing to the ensuing increased particle  localization.
However, in a previous study\cite{saheed} we established that 2D parahydrogen clusters of less than thirty molecules, turning superfluid at a temperature of the order of 1 K, display a rather marked ``supersolid" character (obviously such a definition is necessarily loose, given that we are talking about a finite system); that is, although exchanges of identical molecules are frequent at low temperature, 
concurrently molecules are nonetheless spatially localized and form orderly structures. Thus, owing to their greater rigidity, superfluidity in these finite clusters may be robust against external confinement -- more so than in helium droplets, for instance. Furthermore, it has been recently shown that confinement can actually have an {\it enhancing} effect on the superfluid response of \paraH2 clusters in three dimensions.\cite{ob1}
\\ \indent
In order to investigate quantitatively the effect of confinement on its structure, energetics  and superfluid properties, we have carried out first principle Quantum Monte Carlo simulations of a single, spatially confined 2D \paraH2 cluster at low temperature ($T=0.25$ K).  We studied clusters comprising up to thirty molecules. Confinement in this study is described by means of a simple gaussian well of varying size (typically of the order of a few \AA) and depth (up to 100 K). 
\\ \indent
The comparison of the computed superfluid response of the confined clusters with that of the corresponding free ones   shows that, while as expected superfluidity is suppressed in sufficiently deep wells, nonetheless clusters retain their structure and superfluid response within a rather wide range of confinement parameters, i.e., they are relatively unaffected by the confinement.  
Suppression  of superfluidity takes place gradually as the well is deepened, affecting primarily the largest clusters. On the other hand, clusters with $\sim$ 15 molecules or less remain superfluid, at the low temperature considered here, even when confined  in fairly deep  wells. 
Our physical conclusion is that, even making allowance for the simplicity of the model utilized, superfluid appears to be remarkably resilient in these intriguing few-body systems; phrased alternatively, the quantitative requirements on the strength and size of the confining well may not be particularly  stringent, at least in terms of  ensuring a significant, possibly observable superfluid response of  confined clusters. \\ \indent
The energetics of the system, however, is such that, no matter what values of depth and size of the attractive well one chooses, the distance between adjacent wells must be taken rather large (close to twice  the size of an individual cluster), in order for the low-density cluster crystal  to be energetically stable against the formation of the equilibrium 2D crystal. Consequently, in such a cluster crystal tunnelling of molecules across adjacent sites, necessary to establish a global superfluid response, will be essentially absent, for  practical purposes. Thus, much like others previously explored,\cite{massimo,massimo2}  this approach to stabilize a bulk superfluid phase of \paraH2 appears unlikely to succeed.  
\\ \indent
The remainder of this article is organized as follows: in the next section we describe the microscopic model underlying the calculation and furnish basic computational details; we devote Sec. \ref{results} to a thorough illustration of our results, discussing the physical conclusions in Sec. \ref{conclusions}.

\section{MODEL AND METHODOLOGY}
\label{model}

Our system of interest is modeled as a collection of $N$ {para}hydrogen (\paraH2) molecules, regarded as point particles of spin zero, moving in 2D in the presence of a confining potential. The quantum mechanical many-body Hamiltonian is the given by
\begin{eqnarray}\label{hm}
\hat{H}&=&-\frac{\hbar^2}{2m}\sum_{i=1}^{N}\nabla_{i}^{2}  + \sum_{i<j}v(r_{ij}) + \sum_{i}V({\bf r}_{i})
\end{eqnarray}
where $\hbar^2/2m=12.031$ K\AA$^2$, ${\bf r}_i$ is the position of the $i$th \paraH2 molecule, $r_{ij}\equiv |{\bf r}_{i}-{\bf r}_{j}|$, 
$v$ is the potential describing the interaction of a pair of molecules, while $V$  is the confining potential.  We model $V$ by means of a simple Gaussian well, centered at the origin, namely
\begin{equation}\label{td}
V(r)=-A\ {\rm exp}\biggl(-\frac{r^2}{2\sigma^2}\biggr)
\end{equation}\indent
While a direct experimental realization of such a confining potential may not be straightforward (not to our knowledge anyway), it contains nonetheless all the relevant ingredients to afford qualitative insight into the physics of the system, with a small number of parameters, thereby rendering the interpretation of the results easier.
We use the well-known Silvera-Goldman potential\cite{silveral,note} to describe the pair-wise interaction among \paraH2 molecules.
\\ \indent
The thermodynamics and the structural properties of the above system at low temperature ($T$=0.25 K) have been studied by means of Quantum Monte Carlo simulations based on the continuous-space Worm Algorithm. Since this well-established methodology is extensively described elsewhere, we shall not review it here, referring instead the readers to the original references.\cite{worm,worm2} Details of the calculation are standards, analogous to those employed, for instance, in the simulation of trapped dipolar Bose systems.\cite{trapdip,filinov}.
Besides the energy per molecule, we compute radial density profiles with respect to the centre of the well, as well as global and local  superfluid response of the clusters, using standard estimators for finite systems.\cite{sindzingre89,whaley,mezzacapo08}

\section{RESULTS}  
\label{results}
\subsection{Superfluidity}
\begin{figure}[t]             
\centerline{\includegraphics[scale=0.33]{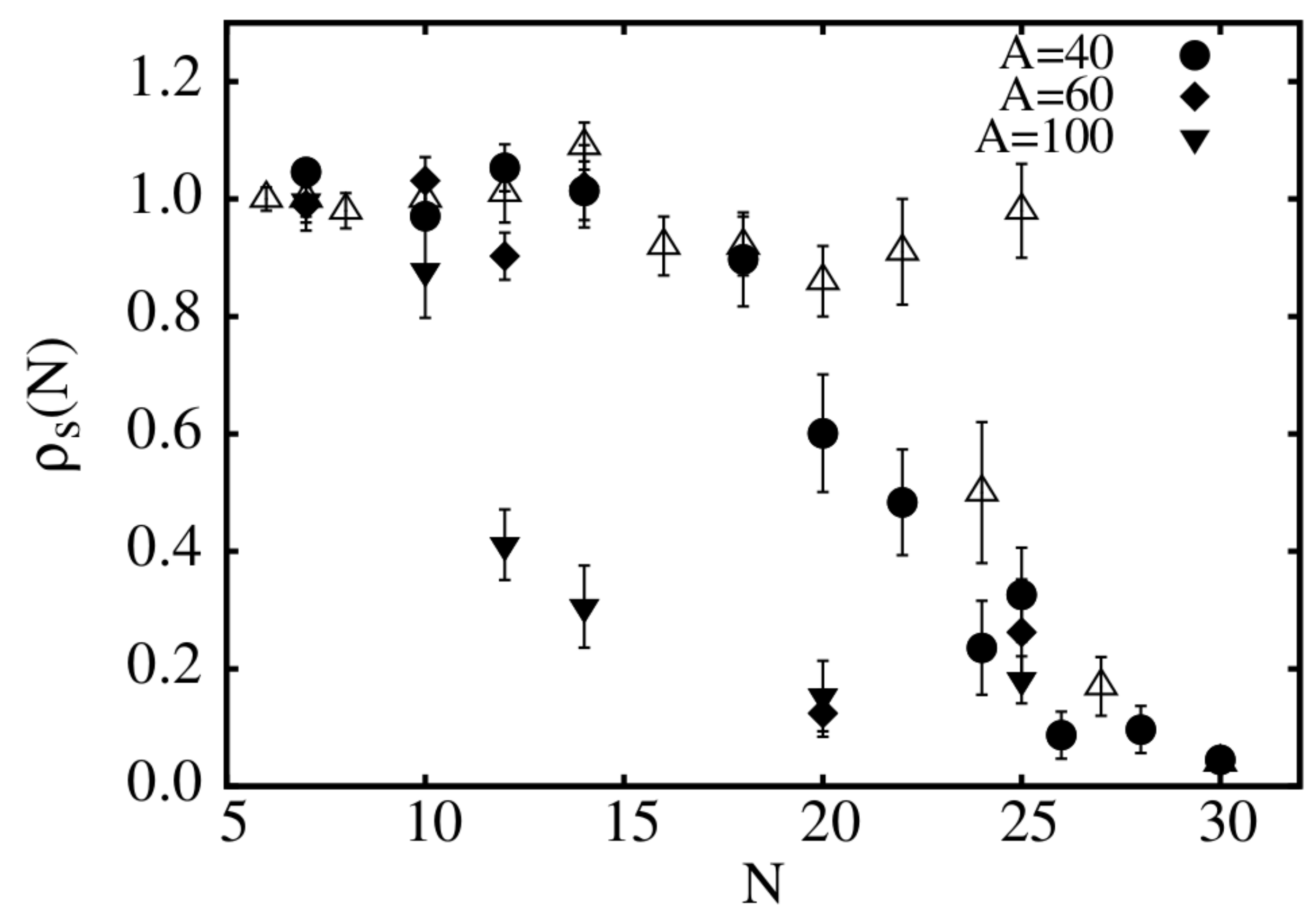}}  
\caption{Superfluid fraction of 2D \paraH2 clusters confined in a Gaussian well of size $\sigma$=3 \AA\ and depth $A$=40 K (circles), $A$=60 K (diamonds) and $A$=100 K (triangles), at a temperature $T$=0.25 K. Open triangles show results for free clusters. When not shown, statistical errors are at the most equal to symbol size.}
\label{sd}
\end{figure} 
We begin by illustrating the results of our study for clusters trapped inside a well of size $\sigma$=3 \AA\ (Eq. \ref{td}), i.e., 
roughly  the radius of the inner shell of the clusters.\cite{saheed}
Figure \ref{sd} shows the superfluid fraction $\rho_{s}(N)$ at a temperature $T$=0.25 K, for clusters comprising up to $N$=30 molecules,  for wells of depth $A$=40, 60 and 100 K respectively.  Also shown are the corresponding results for free clusters (i.e., $A$=0), from Ref. \onlinecite{saheed}. 
\begin{figure}             
{\includegraphics[scale=0.34]{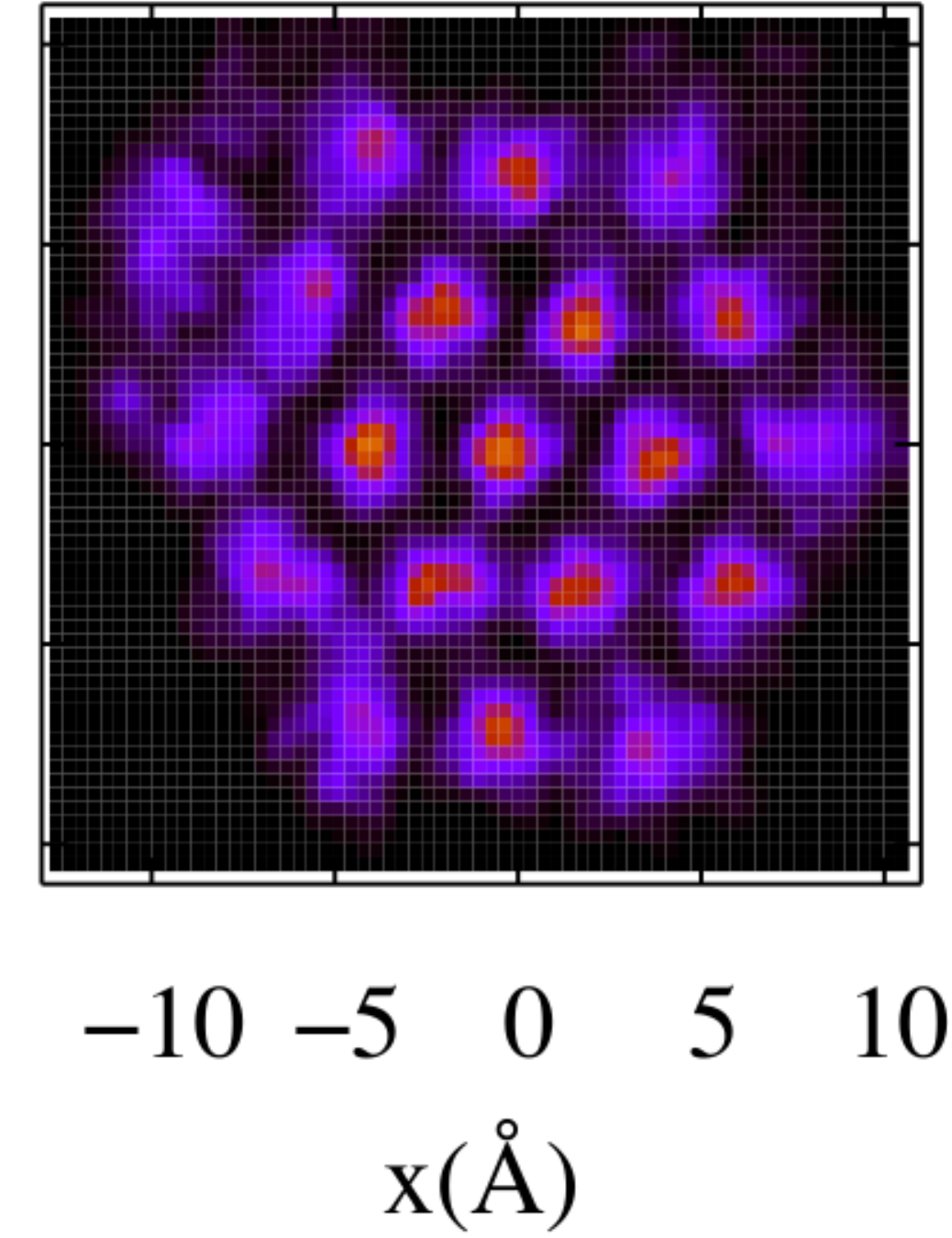}}  
{\includegraphics[scale=0.34]{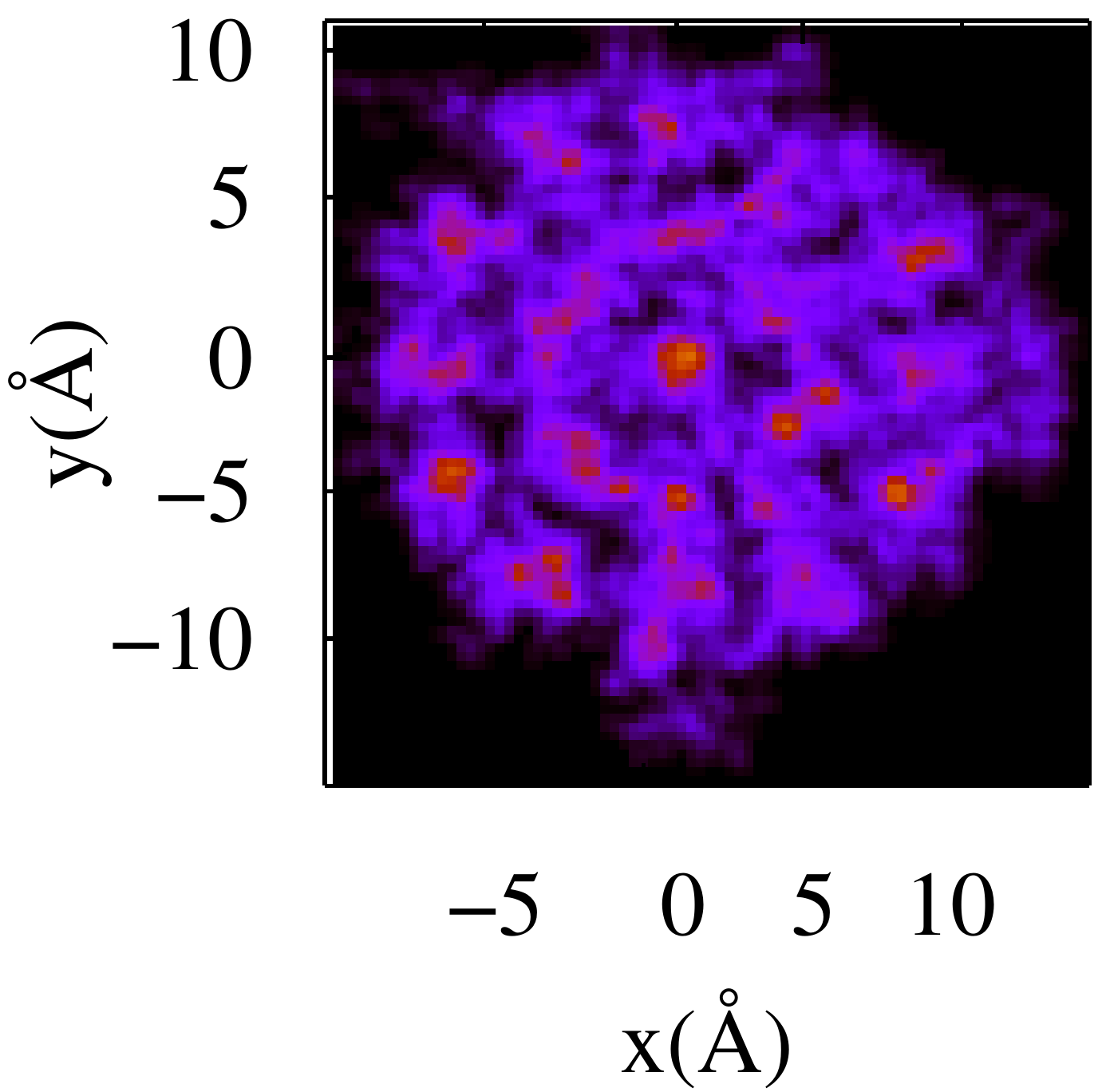}}  
\caption{Configurational snapshots (particle world lines) yielded by a simulation of a cluster with $N$=20 \paraH2 molecules at $T=0.25$ K. {\it Left}: Free cluster. {\it Right}: Cluster confined  inside a gaussian well of depth $A$ = 100 K and size $\sigma$ = 3 \AA. 
Brighter colors correspond to a higher local density.}\label{snap}           
{\includegraphics[scale=0.34]{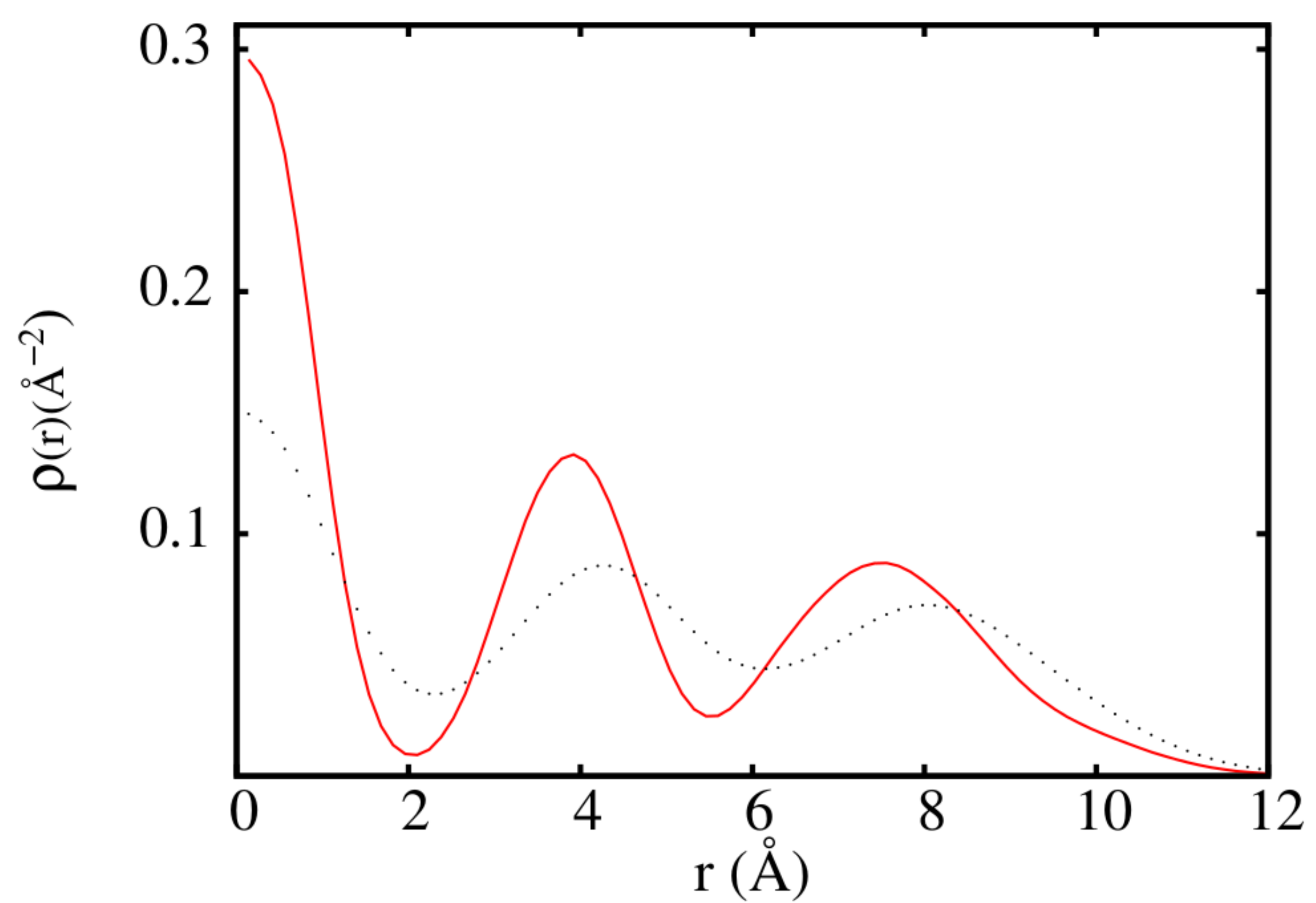}}  
\caption{Radial density profile for a cluster of $N$=20 \paraH2 molecules, confined in a Gaussian well of size $\sigma$=3 \AA\ and of depth $A$=100 K. Dotted line shows the corresponding profile for a free clusters. These profiles are computed at $T$=0.25 K. Statistical errors are not visible on the scale of the figure.}\label{rp}
\end{figure} 
\\ \indent
As discussed therein, a rather sharp demarcation exists for free clusters, in terms of superfluid response. For, those with less than 26 molecules are essentially 100\% superfluid at this low $T$ (with the sole anomaly of $N$=24 for which the superfluid response is approximately one half), whereas the superfluid response is suppressed in larger clusters.
  \\ \indent
For a relatively shallow well ($A\sim 20$ K), the superfluid response of the confined clusters remains close (within $\sim$ 10\%)  to 
that of the free ones. As the depth of the confining well is increased, superfluidity is gradually suppressed, but the smallest clusters, namely those with $N\ \lapx 15$, retain  their superfluid properties (those with $N\lapx 10$ essentially entirely), even for the deepest well considered here, namely with $A$=100 K. 
As shown in Fig. \ref {sd}, clusters whose superfluid response is most significantly affected by confinement are the largest ones, i.e., those with $N\ \gapx18$.\\ \indent
This result may seem counterintuitive, as one might expect confinement to have a more disruptive effect on the superfluidity of smaller clusters.  The reasoning would be that, as the well depth is increased, the molecules in the inner part of the cluster become localized, with the ensuing suppression of quantum exchanges, and thus of superfluidity, which might remain confined to 
the outer region of a larger cluster, where molecules enjoy greater mobility and where the effect of the confining potential is weak,  for a well of size 3 \AA. \\ \indent This is indeed what happens,  as qualitatively shown in Fig. \ref{snap} for a cluster with $N$=20 molecules.  Configurational snapshots for a free cluster (left), and one trapped inside a well of depth $A$=100 K (right),  clearly show a much greater localization of  molecules in the center of the cluster; this is more quantitatively illustrated by the radial density profile, computed with respect to the center of the well, shown in Fig. \ref{rp}.
Hardly any exchanges of \paraH2 molecules take place in the center of such a deep well. Exchanges still occur in the outer shell, but  superfluidity is nonetheless suppressed to statistical noise level in the confined cluster, while it is nearly 100\% in the free cluster. This is consistent with the notion that superfluidity in \paraH2 clusters, which have a strong ``solid-like" structure, crucially hinges on exchanges of molecules across different shells, an effect already noticed in 3D clusters.\cite{mezzacapo08} In the presence of a confining well,  superfluidity is  resilient in smaller clusters, consisting of essentially only one shell,  because molecules are less compressed than in the case of clusters with an additional shell, and therefore enjoy sufficient mobility, even for fairly deep wells ($\sim$ 100 K).
\\ \indent
Structurally, confinement does not induce any significant change, at least in the range of parameters considered here. As shown in the right panel of Fig. \ref{snap}, molecules arrange on a 2D triangular lattice, with no change of structure (for example, number of molecules in the inner ring) for specific values of the confinement parameters, unlike what observed in 2D trapped dipolar bosons, where different solid structures are observed on tightening the confining well.\cite{trapdip}
\begin{figure} [h]         
\centerline{\includegraphics[scale=0.33]{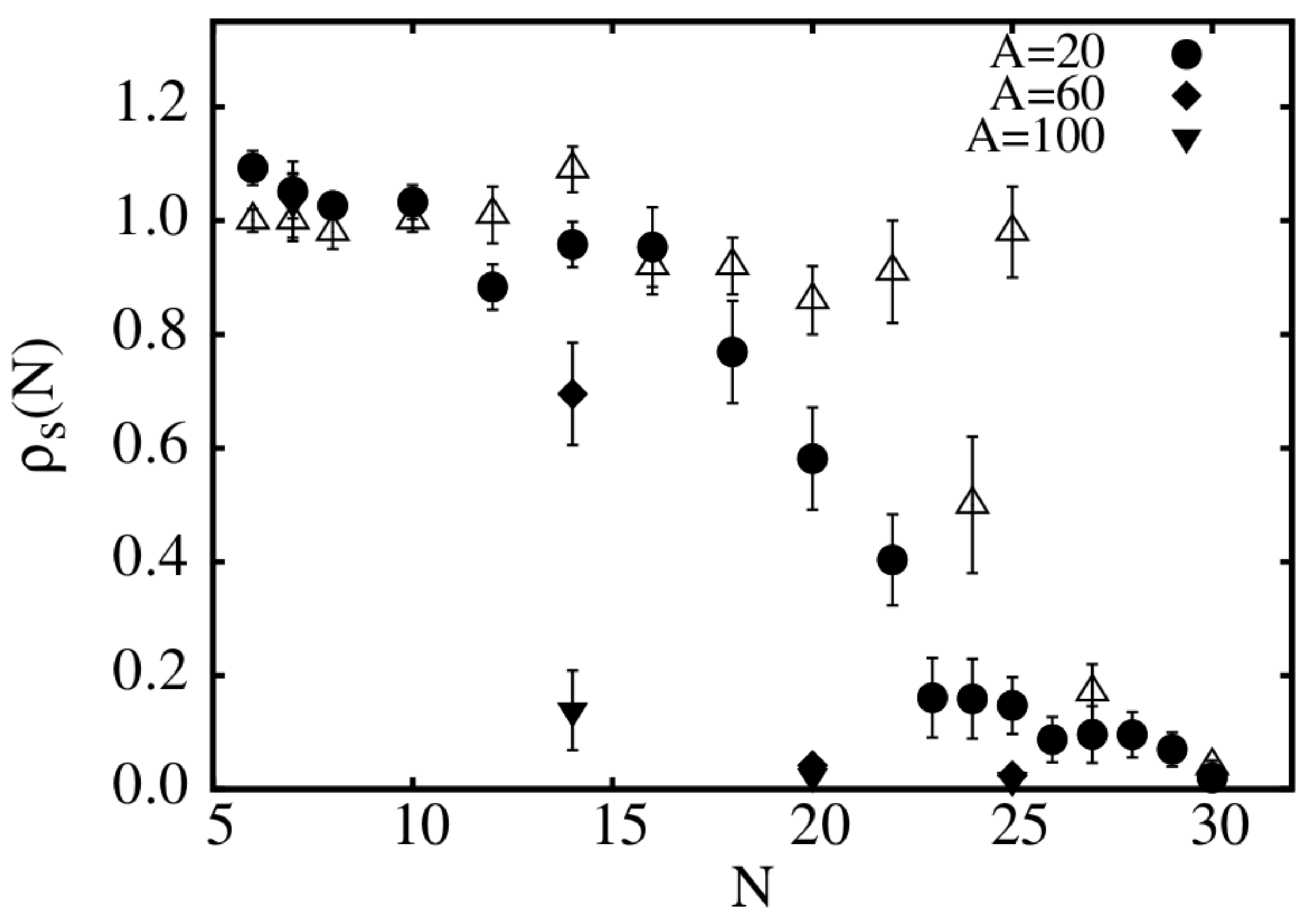}}  
\caption{Superfluid fraction of 2D \paraH2 clusters confined in a Gaussian well of size $\sigma$=6 \AA\ and depth $A$=20 K (circles), $A$=60 K (diamonds) and $A$=100 K (triangles), at a temperature $T$=0.25 K. Open triangles show results for free clusters. When not shown, statistical errors are at the most equal to symbol size.}
\label{ssd}
\end{figure} 
\\ \indent
Fig. \ref{ssd} shows $\rho_S(N)$ for the same clusters as in Fig. \ref {sd}, but for a well of size $\sigma$=6 \AA. The results are qualititatively similar to those obtained for a tighter well, the suppression of superfluidity being more noticeable in this case, especially for clusters comprising between 15 and 20 molecules, for a well of the same depth.  In this case, the confining potential is most rapidly varying rougly between the first and the second shell, for clusters of more than $\sim 15$ molecules, which has a grater suppressing effect for intershell exchanges. The superfluid response of clusters of 15 molecules or less, on the other hand, is relatively unaffected for depths up to 60 K.  The structure of the cluster is affected by the deepening of the confining well in the same way qualitatively 
illustrated in Fig. \ref{snap} for a well of half the radius, i.e., molecules are increasingly more localized.
If the characteristic radius of the well is further increased,  essentially beyond that of the cluster itself, confiment becomes increasingly irrelevant, understandably.
\\
\indent
The main physical conclusion of this part of our study is that the superfluid response of  2D \paraH2 clusters of less than $\lapx$ 20 molecules is quantitatively rather robust against confinement. This is a direct consequence of the ``supersolid" character of these clusters, which renders them less compressible than liquid-like ones,  consequently protecting their main physical properties from the influence of external agents. The basic physics of the superfluid clusters in confinement quantitatively reproduces that of the free clusters.
\subsection{Energetics}
\begin{figure}             
\centerline{\includegraphics[scale=0.33]{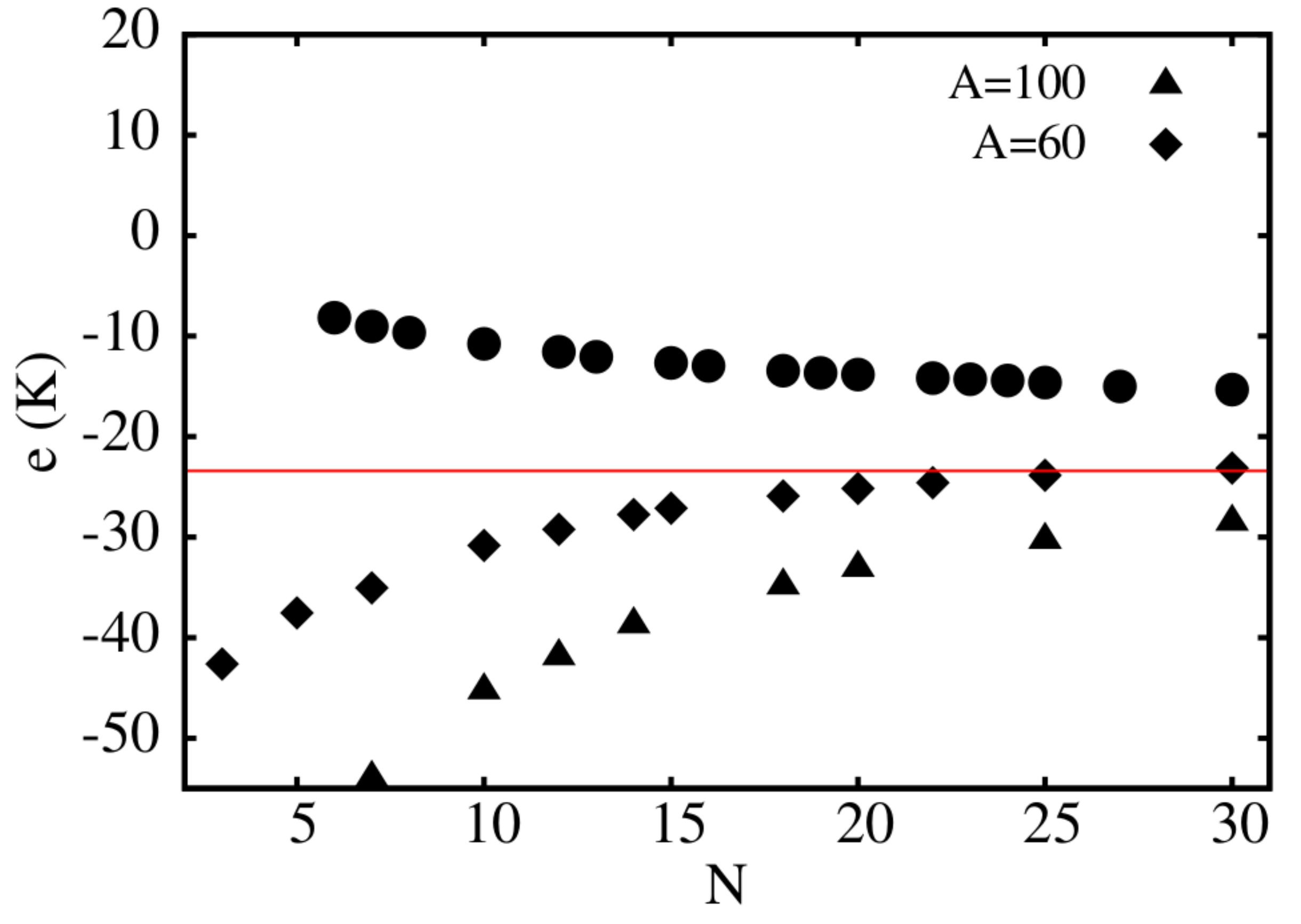}}  
\caption{Energy per \paraH2 molecule for clusters of $N$ molecules ($1\le N\le 30$), trapped in a well of size $\sigma$=3 \AA\  and depth $A$=60 K (diamonds) and $A$=100 K (triangles). Circles show the corresponding values for free clusters (i.e., $A$=0).
Statistical errors are smaller than symbol size. Horizontal line refers to the ground state energy per molecule of bulk \paraH2 in its 2D crystal equilibrium phase.}
\label{ssd1}
\end{figure} 
The energetics of the confined clusters is of interest in view of the possible stabilization of a crystal of 2D clusters, turning superfluid at low temperature, as explained in the Introduction. Fig. \ref{ssd1} shows  the energy per molecule  in a cluster comprising up to thirty molecules, confined in a  Gaussian well of varying amplitude and size $\sigma$=3 \AA. The qualitative behaviour observed in a well of twice the size is the same, all curves being shifted downward. 
\\ \indent
The idea is that of ``pinning" small \paraH2 clusters at the sites of a triangular lattice whose lattice constant $d$ should be of the order of, or not much greater than, the characteristic size of a superfluid cluster, in order to allow for tunnelling of outer shell molecules across adjacent sites. The results shown here and in Ref. \onlinecite{saheed} suggest that $d\sim 20$ \AA. 
\\ \indent
Let us  assume for definiteness a number of molecules per unit cell $N$ equal to 20, yielding a 2D density for the cluster crystal of approximately 0.058 \Am2. This is significantly  less than the ground state equilibrium density of \paraH2 in 2D, equal to\cite{bon04} $\rho_0=0.0667$ \Am2, at which the system is a non-superfluid crystal with one particle per unit cell; the energy per molecule in such a phase is $\epsilon_0=-23.4$ K. In order for the low density cluster crystal phase to be energetically stable against the formation of the ordinary 2D crystal of density $\rho_0$, the energy per \paraH2 molecule should be lower than $\epsilon_0+\Delta$, $\Delta$ being the average potential energy in a lattice of identical wells, of a given lattice constant $d$. This quantity can be easily computed numerically.  \\ \indent 
From Fig. \ref {rp}, we see that the radius of a cluster with 20 molecules is $\sim$ 12 \AA. If the lattice constant $d$ is taken to be 25 \AA, molecules in outer shells would have to tunnel across a distance of $\sim$ 1 \AA. However, for $d$=25 \AA\ and $\sigma$=3 \AA, we have $\Delta\approx -0.107\ A$, consistently shifting the energy per particle of the 2D crystal to a lower value than the energy per particle inside the corresponding well (see Fig. \ref{ssd1}), for any value of the well depth $A$. Thus, the condition of stability of the cluster crystal  is not met, as the system finds energetically more favorable to form its equilibrium 2D crystal (leaving a fraction of the cell empty, as the density is below the equilibrium one). The breaking down of the cluster crystal with the formation of the equilibrium 2D lattice was actually observed in simulation.\\ \indent
If the lattice constant $d$ is taken to be nearly 30 \AA\ (which would entail a rather large tunnelling distance across sites of approximately 6 \AA), then $\Delta \sim -0.073\ A$; in this case,  the cluster crystal becomes energetically favored for $A\sim 100$ K,  but  the superfluidity of the cluster is suppressed in a such a deep well, as shown in Fig. \ref{sd}.  In order to
 make the cluster crystal energetically favorable, for a depth $A$ such that the clusters are still superfluid, the lattice constant $d$ must be taken as large as  36 \AA, making the distance across which molecules would have to tunnel prohibitively large.
\\ \indent
Increasing the width $\sigma$ of the well does not lead to different physics, for a cluster of this many \paraH2 nolecules, because the energy per molecule in the well is shifted downward by an amount roughly equivalent to that by which the magnitude of $\Delta$ is increased. Moreover, the disruptive effect of confinement on the superfluid response is greater for this value of $\sigma$, as a result of which the lattice constant needed to make the cluster crystal thermodynamically stable is again above 30 \AA.
\\ \indent
The energy balance is more favorable for smaller clusters, i.e., $N$=10, whose radius is approximately 8 \AA. In this case, for $\sigma$=3\ \AA, the cluster crystal with $d\sim$ 20 \AA, for which 
$\Delta \sim -0.168\ A$, is  energetically favored for $A\ \gapx 60$ K; it should be noted that clusters of these sizes remain superfluid even for such deep confining wells. Inded, on taking $A\sim 100$ K the cluster 
crystal  is favored over the equilibrium 2D crystal even for $d$ as low as $\sim$ 19 \AA\ (because we are considering molecular tunnelling, a difference of 1 \AA\ is significant).  Tunnelling of molecules across adjacent clusters would involve in this case a distance of 3-4 \AA. Whether that can allow for a superfluid phase at an attainable temperature remains to be established. The main result of this study, however, is that a superfluid cluster crystal phase of \paraH2, if at all attainable, should have a number of molecules per unit cell equal to ten or less.
\section{Conclusions}
\label{conclusions}
The low temperature superfluid response and energetics of small \paraH2 clusters trapped inside a Gaussian confining well 
have been studied by means of Quantum Monte Carlo simulations. The purpose was on the one hand to assess the robustness of the superfluid response predicted for the free clusters,\cite{saheed} on the other that of assessing the possibility of stabilizing a superfluid cluster crystal phase of \paraH2 in 2D, analogous to that observed in simulations for soft core bosons.\cite{saccani,review} \\ \indent
The main physical conclusion is that 2D clusters retain in confinement most of the same physical properties of the free systems, at least within the range of confining parameters explored here. Clearly, the model of confinement adopted here is oversimplified; a more realistic physical model would presumably  describe adsorption sites as impurities around which clusters would coalesce, i.e., with a short-distance repulsion between \paraH2 molecules and the impurity. This may have a suppressing effect on the superfluid response. Also, the effect of foreign substitutional impurities on the superfluidity of the clusters has not been addressed in this study. 
Based on the computed energetics, the stabilization of a superfluid cluster crystal phase seems  possible if clusters are relatively small ($\lapx\ 10$ molecules), for a lattice constant some 20-25\% greater than the characteristic size of the superfluid clusters. This would require \paraH2 molexcules to tunnel across a distance of 3-4 \AA, in order for phase coherence to be established across the whole system. 
Simulation work aimed at ascertaining the existence of such a phase, as well as it possible superfluid properties, is in progress.
\section*{ACKNOWLEDGMENTS}
This work was supported by the Natural Sciences and Engineering Research Council of Canada (NSERC). Computing support from Westgrid is gratefully acknowledged.

\end{document}